\documentclass[aps,prb,twocolumn,amsmath,showpacs,showkeys,floatfix]{revtex4}
\usepackage{epsfig}
\usepackage[english]{babel}
\usepackage{graphicx}
\usepackage{graphics}
\usepackage{amsmath}
\usepackage{url}
\usepackage{amssymb}
\newcommand{\eeq}{\end{equation*}}
\begin{document}
\renewcommand{\figurename}{FIG.}
\newcommand{\vc}[1]{\mathbf{#1}}
\newcommand{\vm}[4]{#1_{#2}\!\setlength{\arraycolsep}{1pt}
                         \renewcommand{\arraystretch}{0.6}
                   \begin{array}{c}{\scriptstyle#3}\\
                                  {\scriptstyle#4}
                   \end{array}\!}
\thispagestyle{empty}

\title{A microscopic modeling of phonon dynamics and charge response in metallic BaBiO$_{3}$}
\author{Claus Falter}
\email[Email to: ]{falter@nwz.uni-muenster.de}
\author{Thomas Bauer}
\author{Thomas Trautmann}
\affiliation{Institut f\"ur Festk\"orpertheorie, Westf\"alische
Wilhelms-Universit\"at,\\ Wilhelm-Klemm-Str.~10, 48149 M\"unster,
Germany}
\date{\today}

\begin{abstract}
We use our recently proposed microscopic modeling in the framework of linear response theory to investigate the complete phonon dispersion, the phonon density of states, certain
phonon-induced electronic charge distributions and charge fluctuations (CF's) for anomalous soft modes of metallic BaBiO$_{3}$ in its simple cubic phase where superconductivity with $T_{c}$ up
to 32 K appears. The theoretical approach already has been applied successfully to the cuprate high-temperature superconductors (HTSC's), simple ionic crystals (NaCl, MgO) and perovskite
oxides (SrTiO$_{3}$, BaTiO$_{3}$). It is well suited for materials with a strong component of ionic binding and especially for "ionic" metals. In particular, the giant phonon anomalies related to the breathing
vibration of the oxygen as found experimentally in superconducting doped Ba$_{0.6}$K$_{0.4}$BiO$_{3}$, resembling those observed in the high $T_{c}$ cuprates, are
investigated.  The origin of these anomalies is explored and attributed to a strong nonlocal coupling of the displaced oxygen ions to CF's of ionic type,
essentially of the Bi6s- and Bi6p orbital. This points to the importance of both of these states at the Fermi energy. Starting from an ab-initio rigid ion model (RIM) we calculate the
effect on the lattice dynamics and charge response of the most important electronic polarization processes in the material, i.\,e. CF's and dipole fluctuations (DF's). Taking into account
these electronic degrees of freedom in linear response theory, we obtain a good agreement with the measured phonon dispersion and in particular with the strong phonon anomalies.
\end{abstract}

\pacs{63.20.Dj, 74.70.-b, 74.25.Kc, 63.20.Kr}

\keywords{high-temperature superconductors,
lattice dynamics, electron-phonon-interaction,
electronic state, density response}

\maketitle

\section{Introduction}
Potassium doped Ba$_{1-x}$K$_{x}$BiO$_{3}$ with a transition temperature $T_{c}$ up to 32 K for $x \approx 0.4$ has been intensively studied with no final resolution
of its high-temperature superconductivity. Different from the cuprate high-temperature superconductors (HTSC's) its bonding is three dimensional and no antiferromagnetic ordering exists
for the insulating parent compound indicating that there are no strong electronic correlations near $x = 0$. Common to both materials, however, is the strong component of ionic
binding favoring long-ranged Coulomb interactions and a related strong nonlocal electron-phonon coupling in terms of localized charge fluctuations (CF's) of the ionic orbitals.
This type of coupling has been shown in the cuprate-HTSC's to lead to the generic phonon anomalies of the high-frequency oxygen bond-stretching modes (OBSM) \cite{Ref01,Ref02,Ref03,Ref04,Ref05}. Recently,
giant phonon anomalies for similar OBSM (oxygen breathing modes) also have been found by inelastic neutron scattering measurements in superconducting Ba$_{0.6}$K$_{0.4}$BiO$_{3}$
\cite{Ref07,Ref08,Ref09}. Thus, while spin-degrees of freedom seem to play no decisive role for superconductivity in Ba-Bi-O the effects of nonlocal coupling of lattice and charge degrees of freedom
in terms of ionic CF's  are still present, likewise as in the cuprate-HTSC's, and provide a coupling channel for pairing in a phonon-mediated mechanism.

Ba$_{1-x}$K$_{x}$BiO$_{3}$ has a complex structural phase diagram \cite{Ref10}. For the insulating composition $0 \leq x < 0.12$ the structure is monoclinic and may be derived from the cubic
perovskite structure by simultaneous octahedral tilting $(t)$ and symmetric oxygen breathing-mode $(b)$ distortions (frozen-in OBSM instability at the $R$-point of the cubic phase).
For $0.12 < x < 0.37$ the structure is still insulating but orthorhombic with a $t$, but no $b$ distortion and for the superconducting, metallic composition $(0.37 < x < 0.53)$ the structure
is simple cubic, i.\,e. no $t$ and $b$ distortion.

It is widely believed that the insulating phases are generated by charge density instabilities associated with breathing and tilting distortions but it has proven difficult to establish
this by first principles calculations. Recent investigations \cite{Ref11} indicate, in contrast to earlier results \cite{Ref12}, that the local-density approximation (LDA) seriously underestimates
the breathing distortions and yields a metallic ground state with no sign of a charge-density-wave instability.

Moreover, first principles linear response calculations of the phonon dispersion of cubic Ba-Bi-O \cite{Ref11} overestimate the high-frequency branches by about $20\,\%$ in comparison
with the experiments. In particular the most strongly renormalized anomalous OBSM, $O^{R}_{B}$, at the $R$-point, which according to the experiments \cite{Ref09} is found at about
10.5 THz has a frequency of 15.7 THz in the calculations presented in Ref. \onlinecite{Ref11}. For the modes with lower frequencies there is a better agreement of the order of $10\,\%$. Thus, linear response
calculations based on LDA-like electronic band structures, using a virtual-crystal approximation seem to underestimate the kinetic part of the charge response,  i.\,e. the electronic
polarizability matrix. In order to investigate this point, we study different cases for the polarizability matrix $\Pi_{\kappa\kappa'}(\vc{q}\,)$ in our microscopic
model approach to extract the most important electronic degrees of freedom for the charge response. The indices $\kappa,\,\kappa'$ denote the orbital degrees of freedom
in an elementary cell of the crystal.

Empirical shell model calculations \cite{Ref07,Ref08,Ref09} incorporating metallic screening by the Lindhard function are also unable to reproduce the downward dispersion and softening of the phonon anomalies,
instead an upward dispersion is calculated.
Finally, calculations in Ref. \onlinecite{Ref13} based on a tight-binding representation of the electronic contribution to the dynamical matrix \cite{Ref14} and an empirical contribution for the short-ranged force
constants fitted to the experimental frequencies lead to a remarkable softening for several modes especially around the $M$- and the $R$-point with strongest
softening for the (1, 1, 0) direction, in disagreement with the experiments where only the OBSM of breathing type are strongly renormalized and the largest softening is along
the (1, 1, 1) direction.

In the present work we calculate within our microscopic modeling of the electronic charge response in linear response theory the complete phonon dispersion, the phonon density of states, the phonon-induced
electronic charge redistribution and the CF's for the OBSM, $O^{X}_{B}$, $O^{M}_{B}$, $O^{R}_{B}$, of simple cubic Ba-Bi-O. In the past, this theoretical approach, which is well suited for systems with a
strong component of ionic binding, has been  applied successfully to the cuprate-HTSC's \cite{Ref01,Ref02,Ref03,Ref04,Ref05,Ref06}, i.\,e., "ionic" metals in which the band crossing the Fermi level has admixture of both
anion and cation orbitals, simple ionic crystals \cite{Ref15} and the perovskite oxides (SrTiO$_{3}$, BaTiO$_{3}$) \cite{Ref16}.

From a general point of view our treatment of the electronic
density response and lattice dynamics in terms of dipole
fluctuations (DF's) and CF's can be considered as a microscopic
(semi-ab initio) implementation of the phenomenological
dipole-shell model or the charge-fluctuation models, respectively.
For a general formulation of phenomenological models for lattice
dynamics that use localized electronic variables as adiabatic
degrees of freedom, see, for example, Ref. \onlinecite{Ref17}.
This formulation covers shell models, bond-charge models and
charge-fluctuation models. While in such an approach the coupling
coefficients are treated simply as empirical fitting parameters
the essential point in our scheme is that all the couplings are
microscopically well defined and can be calculated.

The ionic nature of Ba-Bi-O is described by an ab initio rigid ion model (RIM) leading to a local rigid charge response and electron-phonon interaction (EPI), respectively. The missing nonlocal,
nonrigid part of the electronic density response and EPI is expressed by microscopically  well defined CF's and DF's on the outer shells of the ions. Starting from the RIM
as a reference system the effect
of the nonrigid electronic polarization effects in terms of CF's and DF's on the phonon dispersion and, particularly, the anomalies is calculated and compared with the experiment.

The article is organized as follows. In section 2 we outline the theory and modeling to provide a better reading of the article. Section 3 presents our calculated results of the phonon
dispersion and gives a detailed discussion of the effects of screening by DF's and CF's on the phonon anomalies by comparing the corresponding results with those as obtained from
the RIM as a reference system. Moreover, phonon density of states, the charge redistribution and the CF's for the OBSM of breathing type are presented. Finally, section 4 contains a summary and discussion.

\section{Sketch of the theory and modeling}
In the following a survey of the theory and modeling is presented. A more detailed description can be found in Ref. \onlinecite{Ref01} and in particular in Ref. \onlinecite{Ref15}
where the calculation of the microscopic
coupling parameters is given. The local, rigid part of the electronic charge response and the EPI is approximated by an ab initio RIM taking into account ion-softening in terms
of (static) effective ionic charges and scaling of the short-ranged part of certain pair potentials between the ions to simulate covalence effects in the model. This is done in such a way
that the energy-minimized structure is as close as possible to the experimental one \cite{Ref18}. Structure optimization and energy minimization is very important for a reliable calculation
of the phonon dynamics through the dynamical matrix.

The RIM with the corrections just mentioned serves as an unbiased reference system for the investigation of the effect of the nonrigid electronic polarization processes
on the phonon dynamics. The latter are modeled in form of electronic CF's  on the outer shells of the ions. Especially in the metallic state the CF's dominate the nonlocal
contribution of the charge response and the EPI. In addition DF's are admitted in our approach \cite{Ref05,Ref15}. Thus, the basic variable of our model is the ionic density which is given in the perturbed
state by

\begin{equation}\label{Eq1}
\rho_\alpha(\vc{r},Q_\lambda, \vc{p}_\alpha) =
\rho_\alpha^0(r) + \sum_{\lambda}Q_\lambda \rho_\lambda^\text{CF}(r)
+ \vc{p}_\alpha \cdot
\hat{\vc{r}} \rho_\alpha^\text{D}(r).
\end{equation}

$\rho^{0}_{\alpha}$ is the density of the unperturbed ion, as used
in the RIM, localized at the sublattice $\alpha$ of the crystal
and moving rigidly under displacement. The $Q_{\lambda}$ and
$\rho_{\lambda}^{CF}$ describe the amplitude and the form-factors
of the CF's and the last term in Eq. \eqref{Eq1} represents the
dipolar deformation of an ion $\alpha$ with amplitude (dipole
moment) $\vc{p}_{\alpha}$ and a radial density distribution
$\rho^{D}_{\alpha}$. $\hat{\vc{r}}$ denotes the unit vector in the
direction of $\vc{r}$. The $\rho_{\lambda}^{CF}$ are approximated
by a spherical average of the orbital densities of the ionic
shells calculated in LDA taken self-interaction effects (SIC) into
account. The dipole density $\rho^{D}_{\alpha}$ is obtained from a
modified Sternheimer method in the framework of LDA-SIC
\cite{Ref15}. All SIC calculations are performed for the average
spherical shell in the orbital-averaged form according to Ref.
\onlinecite{Ref19}. For the correlation part of the energy per
electron, $\epsilon$, the parametrization given in Ref.
\onlinecite{Ref19} has been used.

The total energy of the crystal is obtained by assuming that the
density can be approximated by a superposition of overlapping
densities $\rho_{\alpha}$. The $\rho_{\alpha}^{0}$ in Eq.
\eqref{Eq1} are also calculated within LDA-SIC taking environment
effects, via a Watson-sphere potential and the static effective
charges of the ions into account. The Watson-sphere method is only
used for the oxygen ions and the depth of the Watson-sphere
potential is set as the Madelung potential at the corresponding
site. Finally, applying the pair-potential approximation we get
the total energy

\begin{equation}\label{Eq2}
E(R,\zeta) = \sum_{\vc{a},\alpha} E_\alpha^\vc{a}(\zeta)
+\frac{1}{2}\sum_{(\vc{a},\alpha)\neq(\vc{b},\beta)}
\Phi_{\alpha\beta}
\left(\vc{R}^\vc{b}_\beta-\vc{R}^\vc{a}_\alpha,\zeta\right).
\end{equation}

The energy $E$ depends on both, the configuration of the ions $\{R\}$ and the electronic (charge) degrees of freedom (EDF) $\{\zeta\}$ of the charge density, i.\,e.
$\{Q_{\lambda}\}$ and $\{\vc{p}_{\alpha}\}$ in Eq. \eqref{Eq1}. $E^{\vc{a}}_{\alpha}$ are the energies of the single ions. $\vc{a}$, $\vc{b}$ denote the elementary cells and
$\alpha$, $\beta$ the corresponding sublattices. The second term in Eq. \eqref{Eq2} is the interaction energy  of the system expressed in terms of anisotropic pair interactions
$\Phi_{\alpha\beta}$. Both $E^{\vc{a}}_{\alpha}$ and $\Phi_{\alpha\beta}$ in general depend upon $\zeta$ via $\rho_{\alpha}$ in Eq. \eqref{Eq1}.

The pair potentials in Eq. \eqref{Eq2} can be separated into long-ranged Coulomb contributions and short-ranged terms as follows:

\begin{align}\nonumber
\Phi_{\alpha\beta}(\vc{R},\zeta) =& \frac{\mathcal{Z}_\alpha \mathcal{Z}_\beta}{R}
-(\mathcal{Z}_\alpha \vc{p}_\beta + \mathcal{Z}_\beta \vc{p}_\alpha)\cdot\frac{\vc{R}}{R^3}
+\frac{\vc{p}_\alpha\cdot\vc{p}_\beta}{R^3}\\\label{Eq3}
&-3\frac{(\vc{p}_\alpha\cdot\vc{R})(\vc{R}\cdot\vc{p}_\beta)}{R^5}
+ \widetilde{\Phi}_{\alpha\beta}(\vc{R},\zeta),
\end{align}

\begin{align}\nonumber
\widetilde{\Phi}_{\alpha\beta}(\vc{R},\zeta) =& K_\alpha U_\beta(\vc{R},\zeta)
+ K_\beta U_\alpha(\vc{R},\zeta)\\\label{Eq4}
&+ W_{\alpha\beta}(\vc{R},\zeta) + G_{\alpha\beta}(\vc{R},\zeta).
\end{align}

The first term in Eq. \eqref{Eq3} describes the long-ranged ion-ion, the second the dipole-ion and the third and fourth term the dipole-dipole interaction. ${\cal{Z}}_{\alpha}$ and
${\cal{Z}}_{\beta}$ are the variable charges of the ions in case CF's are excited. The latter reduce to the ionic charges for rigid ions. $K_{\alpha}$ and $K_{\beta}$ are the
charges of the ion cores. $\widetilde{\Phi}_{\alpha\beta}$ represents the short-ranged interactions. These can be expressed by the following integrals
\begin{align}\label{Eq5}
U_\alpha(\vc{R},\zeta) &= - \int d^3r \rho_\alpha(\vc{r},\zeta)
\left( \frac{1}{|\vc{r}-\vc{R}|} - \frac{1}{R} - \frac{\vc{r}\cdot\vc{R}}{R^3} \right),
\\\nonumber
W_{\alpha\beta}(\vc{R},\zeta) &= \int d^3r \int d^3r'
\bigl[\rho_\alpha(\vc{r},\zeta) \rho_\beta(\vc{r}',\zeta) \times\\\label{Eq6}
&\times\left(\frac{1}{|\vc{r}-\vc{r}'-\vc{R}|}-\frac{1}{R}-\frac{(\vc{r}+\vc{r}')\cdot\vc{R}}{R^3} \right) \bigr],
\\\nonumber
G_{\alpha\beta}(\vc{R},\zeta) &= \int  d^3r \bigl[\rho_{\alpha\beta}(\vc{r},\zeta)\epsilon(\rho_{\alpha\beta}(\vc{r},\zeta))\\
&-\rho_{\alpha}(\vc{r},\zeta)\epsilon(\rho_{\alpha}(\vc{r},\zeta))\\\nonumber
&-\rho_{\beta}(\vc{r}-\vc{R},\zeta)\epsilon(\rho_{\beta}(\vc{r}-\vc{R},\zeta))\bigr],\label{Eq7}
\end{align}
with

\begin{equation}\label{Eq8}
\rho_{\alpha\beta}(\vc{r},\zeta) = \rho_\alpha(\vc{r},\zeta)+\rho_\beta(\vc{r}-\vc{R},\zeta).
\end{equation}

$K_{\alpha}\,U_{\beta}(\vc{R},\,\zeta)$ yields the short-ranged contribution of the interaction between the core $\alpha$ and the density $\rho_{\beta}$ according to Eq. \eqref{Eq1}.
$W_{\alpha\beta}(\vc{R},\,\zeta)$ represents the short-ranged Coulomb contribution of the interaction of the density $\rho_{\alpha}$ with the density $\rho_{\beta}$ and
$G_{\alpha\beta}(\vc{R},\,\zeta)$ is the sum of the kinetic one-particle- and the exchange-correlation (XC) contribution of the interaction between the two ions \cite{Ref15}. The
short-ranged part of the potentials and the various coupling coefficients are calculated numerically for a set of distances $R$ between the ions. The results so obtained
are then described by an analytical function of the form

\begin{equation}\label{Eq9}
f(R) = \pm \text{exp}\left(\alpha+\beta R+\frac{\gamma}{R}\right).
\end{equation}

$\alpha$, $\beta$, and $\gamma$ in Eq. \eqref{Eq9} are fit parameters.

From the adiabatic condition

\begin{equation}\label{Eq10}
\frac{\partial E(R,\zeta)}{\partial \zeta} = 0
\end{equation}

an expression for the atomic force constants, and accordingly the dynamical matrix in harmonic approximation consistent with linear response theory can be derived:

\begin{align}\label{Eq11}\nonumber
t_{ij}^{\alpha\beta}(\vc{q}) &=
\left[t_{ij}^{\alpha\beta}(\vc{q})\right]_\text{RIM}\\ &-
\frac{1}{\sqrt{M_\alpha M_\beta}} \sum_{\kappa,\kappa'}
\left[B^{\kappa\alpha}_i(\vc{q}) \right]^{*} \left[C^{-1}(\vc{q})
\right]_{\kappa\kappa'} B^{\kappa'\beta}_j(\vc{q}).
\end{align}

The first term on the right hand side denotes the contribution from the RIM. $M_{\alpha}$, $M_{\beta}$ are the masses of the ions and $\vc{q}$ is a wave-vector from
the first Brillouin zone. The quantities $\vc{B}(\vc{q}\,)$ and $C(\vc{q}\,)$ in Eq. \eqref{Eq11} represent the Fourier transforms of the electronic coupling coefficients
as calculated from the energy in Eq. \eqref{Eq2}, and the pair potentials in Eqs. \eqref{Eq3}-\eqref{Eq8}, respectively:

\begin{align}\label{Eq12}
\vc{B}_{\kappa\beta}^{\vc{a}\vc{b}} &= \frac{\partial^2
E(R,\zeta)}{\partial \zeta_\kappa^\vc{a} \partial R_\beta^\vc{b}},
\\\label{Eq13} C_{\kappa\kappa'}^{\vc{a}\vc{b}} &= \frac{\partial^2
E(R,\zeta)}{\partial \zeta_\kappa^\vc{a} \partial
\zeta_{\kappa'}^\vc{b}}.
\end{align}

$\kappa$ denotes the EDF (CF and DF in the present model) in an elementary cell. The $\vc{B}$ coefficients describe the coupling between the EDF and the displaced ions
(bare electron-phonon coupling), and the coefficients $C$ determine the interaction between the EDF. The phonon frequencies, $\omega_{\sigma}(\vc{q}\,)$, and the corresponding
eigenvectors, $\vc{e}^{\,\alpha}(\vc{q}\,\sigma)$, of the modes $(\vc{q}\,\sigma)$ are obtained from the secular equation for the dynamical matrix in Eq. \eqref{Eq11}, i.\,e.

\begin{equation}\label{Eq14}
\sum_{\beta,j} t_{ij}^{\alpha\beta}(\vc{q})e_j^\beta(\vc{q}) =
\omega^2(\vc{q}) e_i^\alpha(\vc{q}).
\end{equation}

The lengthy details of the calculation of $\vc{B}$ and $C$ cannot be reviewed in this paper. They are given in Ref. \onlinecite{Ref15}. In this context we remark that the coupling matrix
$C_{\kappa\kappa'}(\vc{q}\,)$ of the EDF-EDF interaction, whose inverse appears in Eq. \eqref{Eq11} for the dynamical matrix, can be written in matrix notation as

\begin{equation}\label{Eq15}
C = \Pi^{-1} + \widetilde{V}.
\end{equation}

$\Pi^{-1}$ contains the kinetic single particle contribution to
the interaction $C$ and $\widetilde{V}$ the Hartree and
exchange-correlation contribution. $C^{-1}$ needed for the
dynamical matrix and the EPI is closely related to the (linear)
density response function (matrix) and to the inverse dielectric
function (matrix) $\varepsilon^{-1}$, respectively. Comparing with
calculations of the phonon dispersion and the EPI using the linear
response method in form of density functional perturbation theory
within local density approximation (LDA), these calculations
correspond to calculating $\Pi$ and $\widetilde{V}$ in DFT-LDA. On
the oder hand, in our microscopic modeling DFT-LDA-SIC
calculations are performed for the various densities in Eq.
\eqref{Eq1} in order to obtain the coupling coefficients $\vc{B}$
and $\widetilde{V}$. SIC as a correction for a single particle
term is important for contracting in particular localized
orbitals. Written in matrix notation we get for the density
response matrix the relation

\begin{equation}\label{Eq16}
C^{-1} = \Pi(1+\widetilde{V}\Pi)^{-1} \equiv \Pi \varepsilon^{-1},
\hspace{.7cm} \varepsilon = 1 + \widetilde{V}\Pi.
\end{equation}

The CF-CF submatrix of the matrix $\Pi$ can for example approximated from a tight-binding approximation of a single particle electronic  bandstructure. In this case the
electronic polarizability $\Pi$ reads

\begin{align}\nonumber
\Pi_{\kappa\kappa'}&(\vc{q},\omega=0) = -\frac{2}{N}\sum_{n,n',\vc{k}}
\frac{f_{n'}(\vc{k}+\vc{q})
-f_{n}(\vc{k})}{E_{n'}(\vc{k}+\vc{q})-E_{n}(\vc{k})}
\times \\\label{Eq17} &\times \left[C_{\kappa n}^{*}(\vc{k})C_{\kappa
n'}(\vc{k}+\vc{q}) \right] \left[C_{\kappa' n}^{*}(\vc{k})C_{\kappa'
n'}(\vc{k}+\vc{q}) \right]^{*}.
\end{align}

$f$, $E$ and $C$ in Eq. \eqref{Eq17} are the occupation numbers, the single-particle energies and the expansion coefficients of the Bloch-functions in terms of tight-binding functions.

As a measure of the strength of the EPI the self-consistent change of an EDF on an ion, induced by a phonon mode $(\vc{q},\,\sigma)$, can be derived in the form

\begin{equation}\label{Eq18}
\delta\zeta_\kappa^\vc{a}(\vc{q}\sigma) = \left[-\sum_\alpha
\vc{X}^{\kappa\alpha}(\vc{q})\vc{u}_\alpha(\vc{q}\sigma)\right]
e^{i\vc{q}\vc{R}_\kappa^\vc{a}}
\equiv \delta\zeta_\kappa(\vc{q}\sigma)e^{i\vc{q}\vc{R}^\vc{a}},
\end{equation}

with the displacement of the ions

\begin{equation}\label{Eq19}
\vc{u}_\alpha^{\vc{a}}(\vc{q}\sigma) =
\left(\frac{\hbar}{2M_\alpha\omega_\sigma(\vc{q})}
\right)^{1/2}\vc{e}^\alpha(\vc{q}\sigma)e^{i\vc{q}\vc{R}^\vc{a}}
\equiv \vc{u}_\alpha(\vc{q}\sigma)e^{i\vc{q}\vc{R}^\vc{a}}.
\end{equation}

The quantity $\vc{X}$ in Eq. \eqref{Eq18}, i.\,e. the self-consistent response per unit displacement of the EDF is calculated in linear response theory as

\begin{equation}\label{Eq20}
\vc{X}(\vc{q}) = \Pi(\vc{q})\varepsilon^{-1}(\vc{q})\vc{B}(\vc{q}) =
C^{-1}(\vc{q})\vc{B}(\vc{q}).
\end{equation}

Another measure  of the EPI for a certain phonon mode $(\vc{q}\,\sigma)$ is provided by the change of the self-consistent potential in the crystal felt by an electron
at a space point $\vc{r}$, i.\,e. $\delta V_\text{eff}(\vc{r},\,\vc{q}\,\sigma)$. Averaging this quantity with the corresponding density form factor $\rho_{\kappa}(\vc{r} - \vc{R}^{\vc{a}}_{\kappa})$
of the EDF located at $\vc{R}^{\vc{a}}_{\kappa}$, we obtain

\begin{equation}\label{Eq21}
\delta V_\kappa^\vc{a}(\vc{q}\sigma) = \int dV
\rho_\kappa(\vc{r}-\vc{R}_\kappa^\vc{a}) \delta
V_\text{eff}(\vc{r},\vc{q}\sigma).
\end{equation}

This gives an orbital resolved measure for the strength of the EPI
in the mode $(\vc{q}\,\sigma)$ mediated by the EDF considered. For
an expression of $\delta V_{\kappa}^{\vc{a}}(\vc{q}\,\sigma)$ in
terms of coupling coefficients in Eqs. \eqref{Eq12} and
\eqref{Eq13}, see Ref. \cite{Ref20}.

\section{Results and Discussion}
\subsection{Results within the reference System (RIM)}
\begin{figure*}
\includegraphics[]{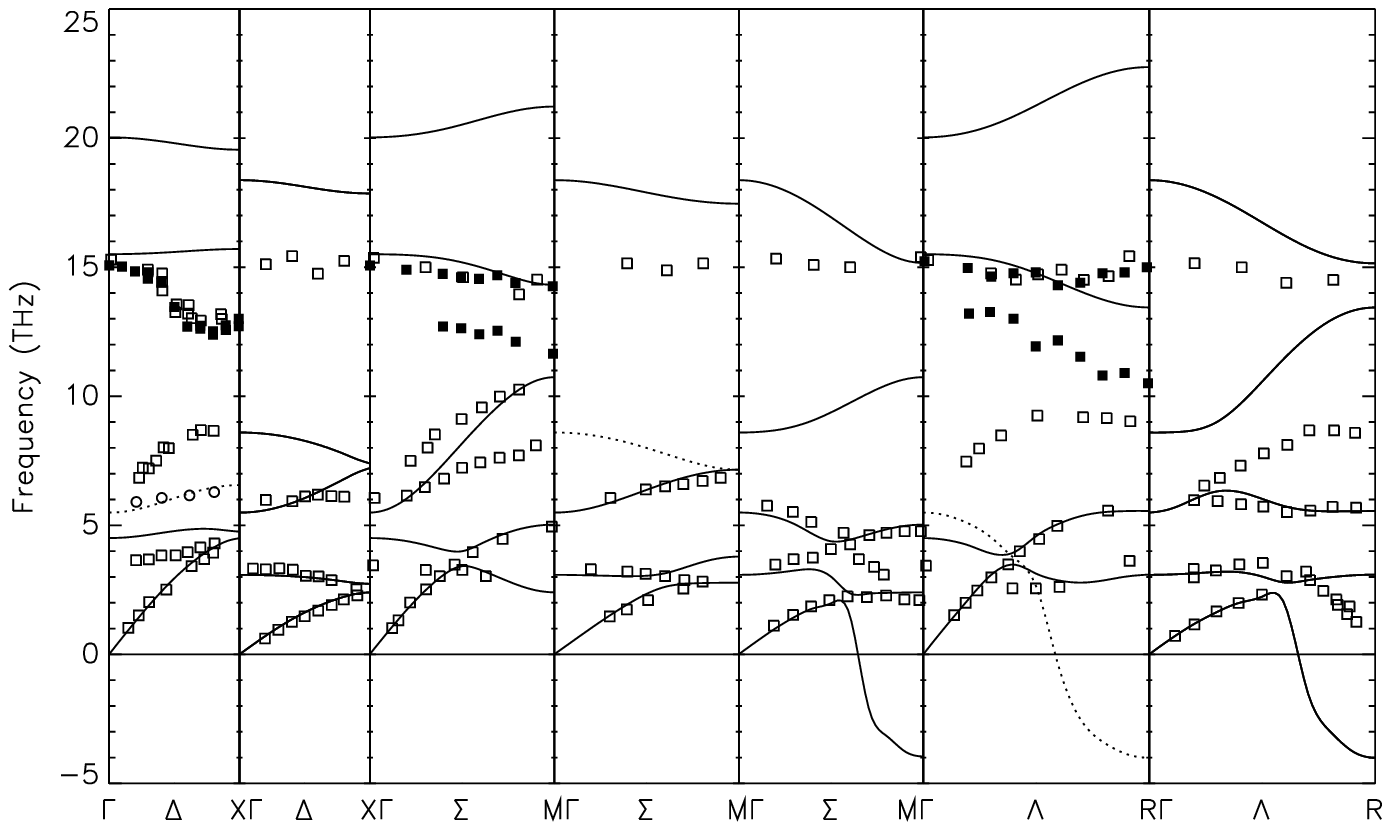}
\caption{
Calculated phonon dispersion of Ba-Bi-O in the cubic perovskite structure in the main symmetry directions $\Delta \sim (0,\,0,\,1)$, $\Sigma \sim (1,\,1,\,0)$ and
$\Lambda \sim (1,\,1,\,1)$ for the ionic reference model with covalent corrections (RIM, model 1). The open symbols represent the experimental results for cubic Ba$_{0.6}$ K$_{0.4}$
BiO$_{3}$ \cite{Ref07}. The full squares are new experimental results of the same material \cite{Ref09}. The arrangement of the panels from left to right according to different irreducible
representations is as follows: $|\Delta_{1}\,(\text{---}),\,\Delta_{2}\,(\cdots)|$ $\Delta_{3}\,(\text{---})|\Sigma_{1}\,(\text{---})|\Sigma_{2}\,(\text{---}),$
$\Sigma_{3}\,(\cdots)|$ $|\Sigma_{4}\,(\text{---})|\Lambda_{1}\,(\text{---})$, $\Lambda_{2}\,(\cdots)|\Lambda_{3}\,(\text{---})|$. The measured anomalous soft
oxygen-breathing vibrations at the $X$, $M$ and $R$ point show the following frequencies: $O_{B}^{X} \approx 12.8\,\text{THz}$, $O^{M}_{B} \approx 11.8\,\text{THz}$ and
$O^{R}_{B} \approx 10.5\,\text{THz}$, respectively. Imaginary frequencies of unstable modes are represented as negative numbers.
}\label{fig01}
\end{figure*}

The rigid, local contribution of the electronic charge response and EPI is approximated by an ab initio RIM with corrections for covalence effects in terms of effective ionic
charges (ion softening) and scaling of the short-ranged pair potential between the Bi and oxygen ion. In this way we obtain a suitable reference system, see Fig. \ref{fig01} for the
calculated dispersion, which subsequently allows for the investigation of the characteristic nonrigid screening effects in terms of CF's and DF's on the phonon dispersion and charge response.

The covalent corrections are important because a calculation of the phonon dispersion of cubic Ba-Bi-O with nominal charges leads to a significant overestimation of the width of the phonon
spectrum. Moreover, various unstable modes down to -12 THz emerge if imaginary frequencies of unstable modes are represented as negative numbers. The mode with the highest frequency
($\approx$ 35 THz) in such a model is the oxygen-breathing mode, $O^{R}_{B}$, at the $R$ point which, on the other hand, according to recent experiments \cite{Ref09} is anomalously soft
($\approx$ 10.5 THz), see Fig. \ref{fig01}. Similar results are found in calculations of the phonon dispersion of cuprate high-temperature superconductors when nominal ionic changes are
used \cite{Ref01,Ref21}. In our calculation we find a set of effective (static) ionic charges (Ba 1.5+,  Bi 2.7+, O 1.4$\text{--}$) and a covalent scaling of the Bi-O pair potential which leads to an energy minimized
structure in full agreement with the experimental one for cubic Ba-Bi-O ($a = 4.2742\,\AA$) \cite{Ref10}. As a general rule, partial covalence reduces the amplitude of the static ionic charges
in mixed ionic-covalent compounds like Ba-Bi-O, because the charge  transfer from the cations to the anions is not complete as in the entirely ionic case.

The phonon dispersion based on this set of reduced ionic charges
is displayed in Fig. \ref{fig01} (model 1). Potassium doping is
taken into account using a virtual-mass approximation and
implicitly via the reduced static charges. The phonon dispersion
calculated with such a reference model means a large improvement
compared to the model with nominal charges (not shown). In
particular, the width of the spectrum is considerably reduced and
compares well with the experiments if the missing renormalization
related to the nonrigid charge response (CF's and DF's) is taken
into account, see subsection \ref{subsecB}. Moreover, unstable
modes only appear at the $R$ and the $M$ point. The mode at $R$ is
triply degenerate and represents rotations around any cubic axis.
The rotation is opposite in neighbouring elementary cells in all
cubic directions. The other unstable mode at $M$ is similar to the
most unstable $R$ point mode, with the exception that the rotation
of the octahedra is in the same sense in adjacent cells along the
$z$-axis. Note in this context that, as mentioned in the
introduction, the structural phase transitions in K-doped Ba-Bi-O
are characterized by rotations (tilts) of the BiO$_{6}$ octahedra
around several distinct axis \cite{Ref10} and in cubic
Ba$_{0.6}$K$_{0.4}$BiO$_{3}$ the frequency of the rotational $R$
point mode is extremely low (Fig. \ref{fig01}), consistent with
the tendency of soft rotational modes at $R$ in harmonic theory.
Thus, our results within the harmonic approximation highlight the
importance of the ionic forces for the tilt instabilities and
indicate that anharmonic contributions should stabilize these
modes. Frozen-phonon calculations support this argument because
they predict highly anharmonic potential behaviour for the tilting
of the BiO$_{6}$ octahedra corresponding to the $R$-point
\cite{Ref11,Ref22}.

\begin{figure}
\includegraphics[]{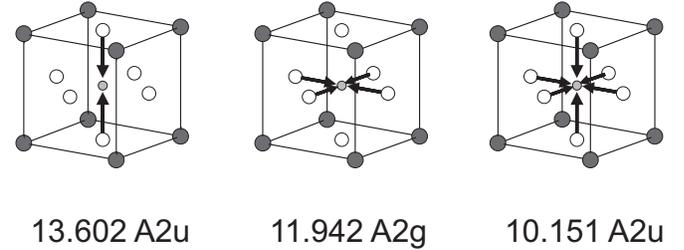}
\caption{Displacement patterns of the anomalous oxygen-breathing modes $O_{B}^{X}$, $O_{B}^{M}$ and $O_{B}^{R}$ from left to right for model 3.
Frequencies are given in units of THz.}
\label{fig02}
\end{figure}

In SrTiO$_{3}$ the rotational modes at $R$ also are found most unstable in our calculations \cite{Ref16} corresponding to an antiferrodistortive transition in agreement with the experiment while
our calculation for BaTiO$_{3}$ find the ferroelectric mode at $\Gamma$ as the most unstable one again in accordance with the experimental situation. Interestingly, we also
obtain in case of the HTSC La$_{2}$CuO$_{4}$ \cite{Ref18,Ref20} one partially unstable branch with the tilt mode at the $X$-point. Freezing-in of this distortion points correctly to the experimentally
observed structural phase transition from the high-temperature tetragonal to the low-temperature orthorhombic structure. Again this rotational instability is already present in the RIM and thus
also driven by the long-ranged ionic interactions likewise as in Ba-Bi-O and Sr-Ti-O. In contrast to the situation in Ba-Ti-O the strongly polar ferroelectric TO mode at
$\Gamma$, where the oxygen anions vibrate coherently in the opposite direction to the Bi ion, is stable in Ba-Bi-O. We get 8.6 THz in the RIM with effective ionic charges and 9.7 THz
 in the RIM with nominal ones. Not only the frequency of the ferroelectric mode is significantly reduced when passing from the model with nominal ionic charges to the model with effective charges
 but also the LO-TO mode splittings.
Inspection of Fig. \ref{fig01} with respect to the oxygen-breathing modes, displayed in Fig. \ref{fig02}, we find that instead of being soft as in the experiment the latter have the highest frequency at the $X\,(O^{X}_{B})$,
$M\,(O^{M}_{B})$ and $R\,(O^{R}_{B})$ point, respectively. Consequently, the characteristic downward dispersion of the anomalous phonon-branches cannot be described at all within the RIM, i.\,e.,
by a rigid, local charge response. In the following we will show that the anomalous dispersion and phonon softening characteristic for cubic, metallic Ba-Bi-O is essentially related
to strong nonlocal, nonrigid EPI effects mediated by CF's of the Bi6s and Bi6p orbitals, pointing to a substantial character also of Bi6p electrons in the electronic state near the
Fermi energy.

\subsection{The effects of a nonrigid, nonlocal charge response via CF's and DF's}\label{subsecB}
Starting from the reference systems (RIM) of the last subsection DF's and CF's as nonrigid electronic degrees of freedom are additionally allowed for
the modeling. This means that on each ion the electrons can redistribute under atomic displacements in such a way that dipole and charge fluctuations are induced on that ion in order to minimize the
energy.

\begin{figure*}
\includegraphics[]{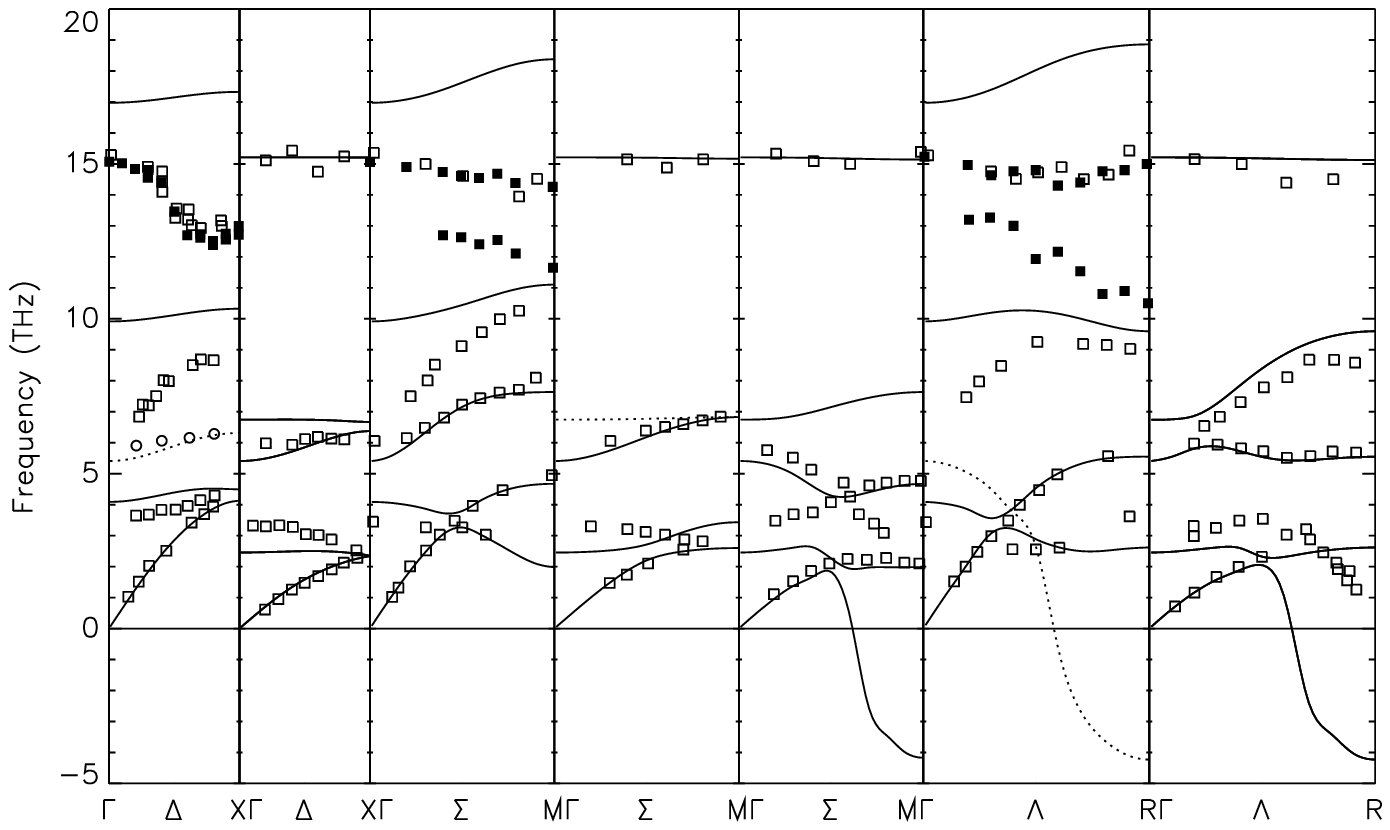}
\caption{Same as in Fig. \ref{fig01} with the calculated results from the model described in the text where additionally to the RIM (model 1) dipole fluctuations are taken into account (model 2).}
\label{fig03}
\end{figure*}

In Fig. \ref{fig03} we display our calculated results of the phonon dispersion taking additionally DF's into account (model 2). In this calculation the ab initio values of the dipole polarizability $\alpha$
as calculated from the Sternheimer method in the framework of DFT-LDA-SIC \cite{Ref15} for the corresponding single ions are reduced by $50\,\%$ which leads to a better
agreement of the phonon dispersion with the experiment. Taking the ab initio values for $\alpha$ as calculated for the single ions with the Sternheimer method overestimates the
dipole polarization in the crystalline environment and leads to TO-frequencies being to low as compared to the experiment. These findings also hold true for our calculations of dipole screening
in the cuprate based HTSC's \cite{Ref05,Ref23} where in addition the dipole polarizability is found to be very anisotropic dominating along the (ionic) $c$-direction.

Comparing Fig. \ref{fig03} with Fig. \ref{fig01} we find that the width of spectrum improves considerably and no additional unstable branches emerge. Moreover, the LO-TO splittings are further reduced
and the TO-modes are well described. In particular, the ferroelectronic TO-mode at $\Gamma$ is reduced from 8.6 to 6.7 THz and is now in quite a good agreement with experiment \cite{Ref07,Ref08}.
However, the anomalous branches with the oxygen-breathing modes at $X$, $M$ and $R$ have an increasing dispersion when propagating from the $\Gamma$ point towards
the zone boundary and show no sign of softening. Likewise as in the RIM, these modes have the highest frequencies in the spectrum.

\begin{figure*}
\includegraphics[]{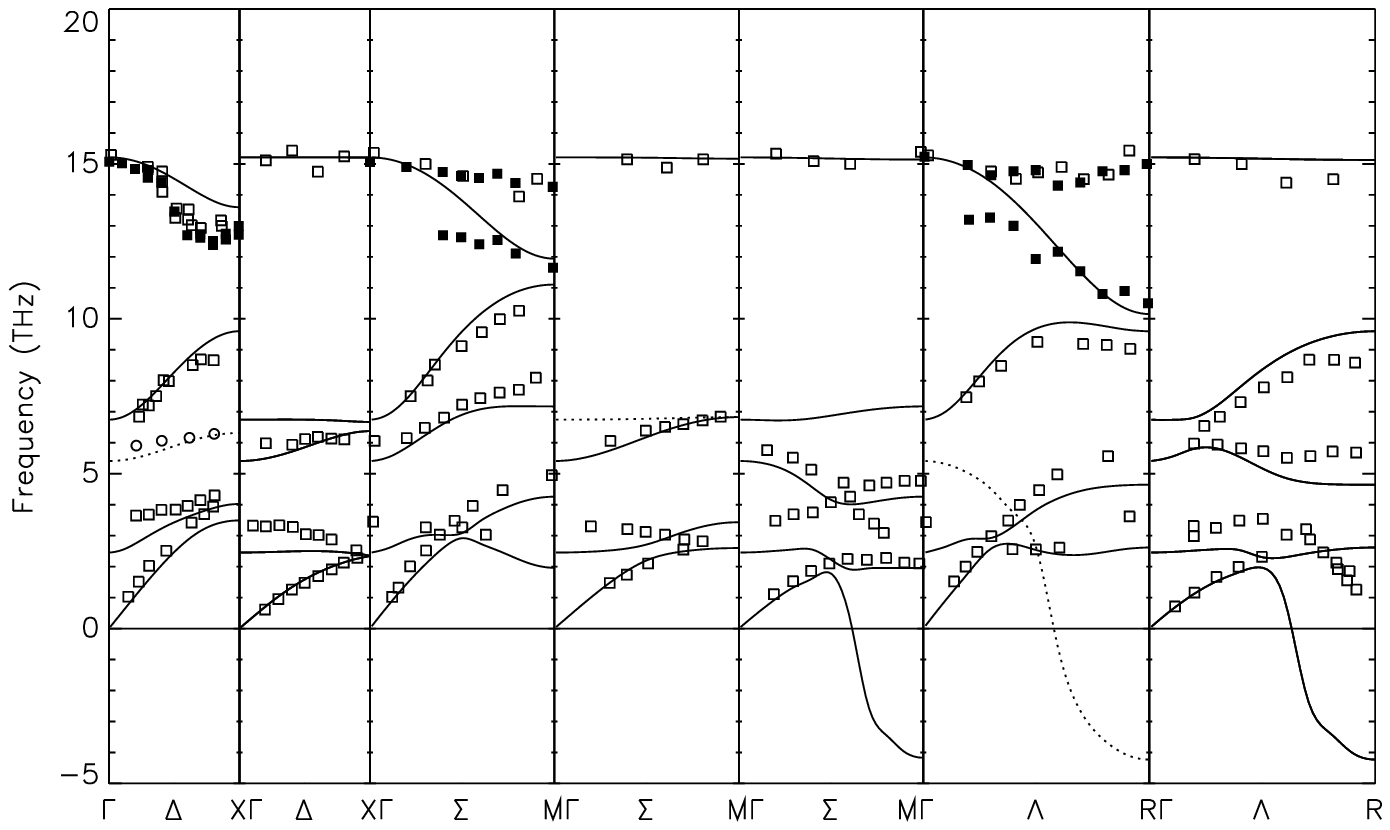}
\caption{Same as in Fig. \ref{fig01} with the calculated results from the model described in the text where additionally to the RIM dipole- and charge fluctuations are included (model 3).}
\label{fig04}
\end{figure*}

Next, we discuss the influence of the ionic CF's on the phonon dispersion. The results of a calculation where additionally to DF's (model 2) CF's are included are shown in Fig. \ref{fig04} (model 3). Comparing the calculated
results with the experiments now a good agreement is achieved. This is particularly true for the phonon anomalies which cannot be described well by other approaches up to now as
stated in the introduction. First principle calculations \cite{Ref11} within DFT-LDA for Ba-Bi-O taking K-doping into account by a virtual-crystal and virtual-mass
approximation report that the breathing distortions are seriously underestimated in the LDA. These calculations yield for the $O_{B}^{R}$ mode a frequency of 15.7 THz compared to 10.5 THz
in the experiment \cite{Ref09}. We argue in the following that a reason for the high frequency in this calculation is the electronic polarizability $\Pi_{\kappa\kappa'}(\vc{q}\,)$ from
Eq. \eqref{Eq17}, i.\,e. the kinetic single particle part of the charge response, which seems underestimated in LDA-type calculations as in Ref. \onlinecite{Ref11} or Ref. \onlinecite{Ref24}, respectively.

In order to investigate this question in our approach, $\Pi_{\kappa\kappa'}(\vc{q}\,)$ is modeled in an appropriate way to find out which electronic degrees of freedom
are most important for the phonon anomalies in Ba-Bi-O. So, we express $\Pi_{\kappa\kappa'}(\vc{q}\,)$ in Fourier-transformed (parametrized) form

\begin{equation}\label{Eq22}
\Pi_{\kappa\kappa'}(\vc{q}) = \sum_{\vc{a}} \vm{\Pi}{}{\vc{0}\,\vc{a}}{\kappa\,\kappa'}
\,\text{e}^{i\vc{q}\left(\vc{R}_{\kappa'}^\vc{a} - \vc{R}_\kappa \right)}.
\end{equation}

In the metallic phase the electronic partial density of states (PDOS) at the Fermi level $Z_{\kappa}(\varepsilon_{F})$ is related to the polarizability matrix in the
long wave length-limit \cite{Ref01,Ref03} according to

\begin{equation}\label{Eq23}
\sum_{\kappa'} \Pi_{\kappa\kappa'}(\vc{q}\rightarrow\vc{0}) =
Z_\kappa(\varepsilon_\text{F}),
\end{equation}

and the total density of states at energy $\varepsilon$ is given by

\begin{equation}\label{Eq24}
Z(\varepsilon) = \sum_{\kappa} Z_\kappa(\varepsilon)
\end{equation}

In our calculations for the metallic phase of the cuprate HTSC's
we have found \cite{Ref03,Ref05,Ref06} that neglecting the
$\vc{q}$-dependence in the polarizability and taking into account
the diagonal on-site parameters
$\vm{\Pi}{}{\vc{0}\,\vc{0}}{\kappa\,\kappa}$ alone already
provides a good approximation for the calculation of the phonon
dispersion, in particular for the anomalous generic OBSM in these
materials. This means that the parameters
$\vm{\Pi}{}{\vc{0}\,\vc{0}}{\kappa\,\kappa}$ are approximated by
the corresponding PDOS. However, such a diagonal approximation
does not hold for the insulating state of a material. Due to the
charge gap in the spectrum off-site elements
$\vm{\Pi}{}{\vc{0}\,\vc{a}}{\kappa\,\kappa'}$ necessarily must
occur and interfere in such a way that the following
compressibility sum rules for the kinetic single particle part of
the charge response are satisfied \cite{Ref01,Ref03,Ref06}.

\begin{equation}\label{Eq25}
\sum_{\kappa'} \Pi_{\kappa\kappa'}(\vc{q}\rightarrow\vc{0}) =
\mathcal{O}(q),
\end{equation}

\begin{equation}\label{Eq26}
\sum_{\kappa\kappa'} \Pi_{\kappa\kappa'}(\vc{q}\rightarrow\vc{0}) =
\mathcal{O}(q^2).
\end{equation}

\begin{figure*}
\includegraphics[]{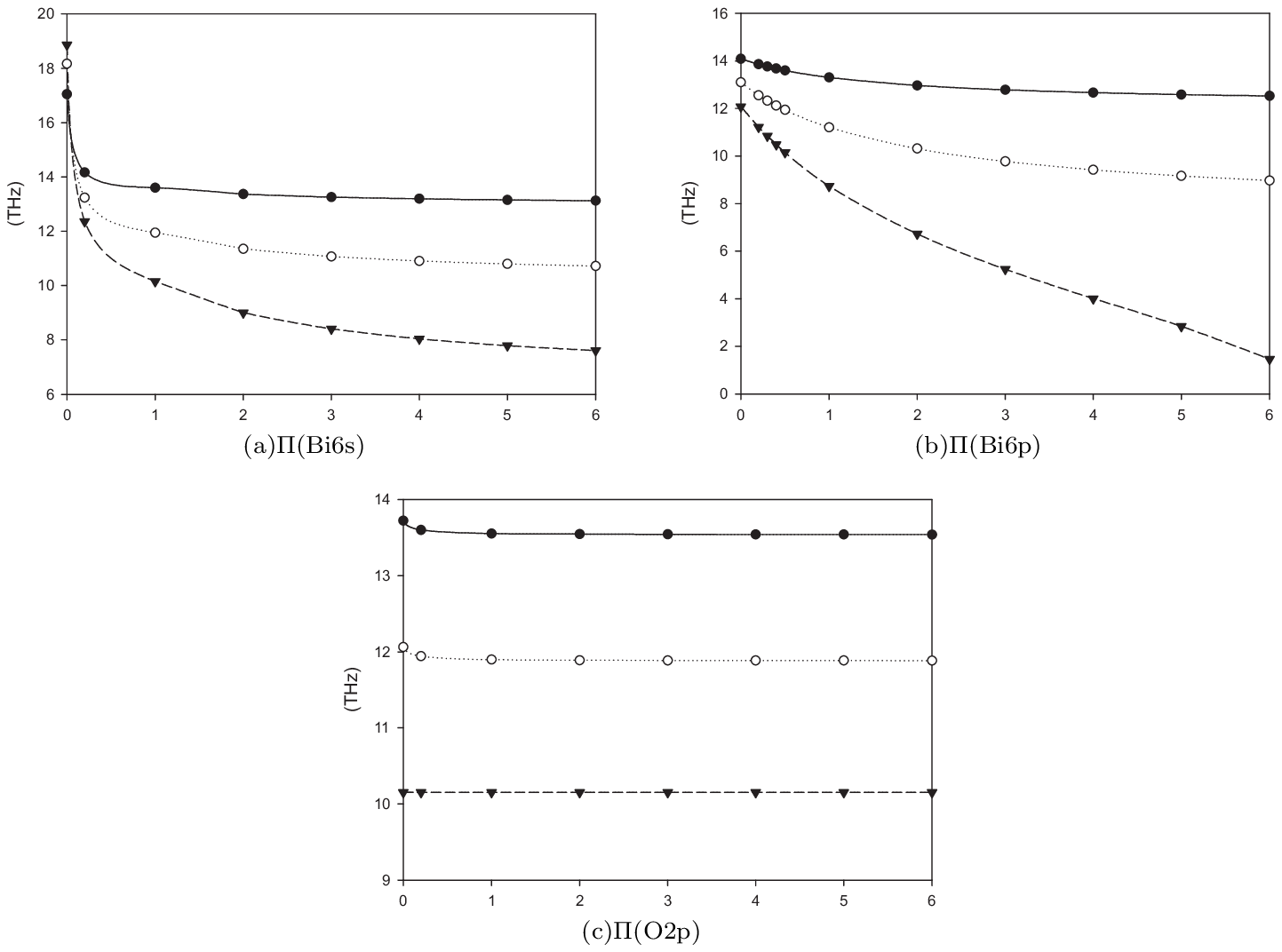}
\caption{Calculated effect of the Bi6s-, Bi6p and O2p charge fluctuations on the softening of the frequency of the anomalous oxygen bond-stretching (breathing) modes
$O_{B}^{X}$, $O_{B}^{M}$ and $O^{R}_{B}$, respectively, for varying polarizabilities $\Pi$(Bi6s), $\Pi$(Bi6p) and $\Pi$(O2p). $\Pi$ is given in eV$^{-1}$.
$X$-point (full line), $M$-point (dotted line), $R$-point (broken line).}
\label{fig05}
\end{figure*}

\begin{figure}
\includegraphics[]{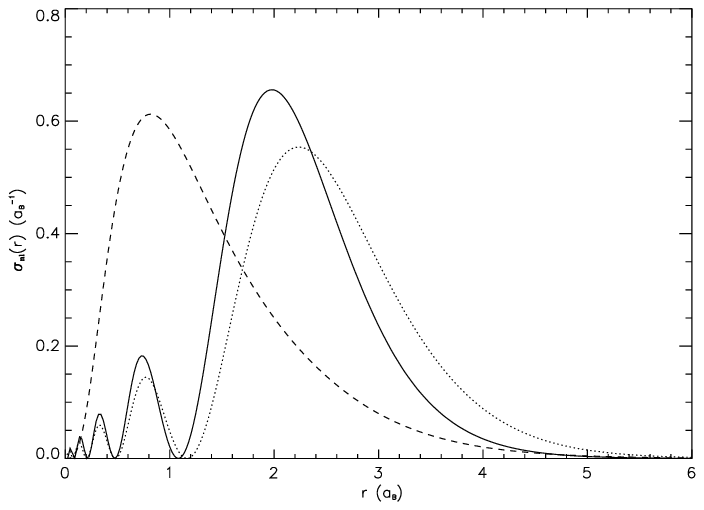}
\caption{Charge density form factors (times $4\,\pi\,r^{2}$) for the Bi$^{2.7+}$ and O$^{1.4-}$ ion. $r$ radial distance in units of $a_{B}$. Bi6s (full curve); Bi6p (dotted curve); O2p
(broken curve).}
\label{fig06}
\end{figure}

In Fig. \ref{fig05} we display the calculated effect of the Bi6s-,
Bi6p- and O2p charge fluctuations for the softening of the
frequency of the anomalous OBSM $O^{X}_{B}$, $O^{M}_{B}$ and
$O^{R}_{B}$, respectively by varying the diagonal on-site
parameters $\Pi\,\text{(Bi6s)} =
Z_{\text{Bi6s}}(\varepsilon_{F})$, $\Pi\,\text{(Bi6p)} =
Z_{\text{Bi6p}}(\varepsilon_{F})$ and $\Pi(\text{O2p}) =
Z_{\text{O2p}}(\varepsilon_{F})$. We find that the CF's in the
Bi6s- and Bi6p orbitals both are important for the phonon
softening, while the dependence on the O2p CF's is very weak. The
values actually chosen for the calculation of the phonon
dispersion in model 3 are $\Pi(\text{Bi6s}) =
1.00\,\text{eV}^{-1}$, $\Pi(\text{Bi6p}) = 0.5\,\text{eV}^{-1}$,
$\Pi(\text{O2p}) = 0.2\,\text{eV}^{-1}$. The latter value, to a
large extent irrelevant for the softening, was taken to optimize
the phonon frequencies for modes where exclusively O2p CF's are
excited. The remarkable softening with increasing
$Z_{\text{Bi6p}}(\varepsilon_{F})$ indicates that the band
crossing the Fermi energy consists not only of the Bi6s and O2p
electrons but also has a substantial admixture of Bi6p electrons.
This is also concluded from electronic band structure
calculations, see Ref. \onlinecite{Ref11} and references therein,
and is consistent with the static effective charge for Bi in our
model pointing out a partial occupation of the Bi6p orbital. The
latter is the most delocalized  one of the the orbitals used in
our calculation (see Fig. \ref{fig06}) and consequently very
effective for screening and softening of the oxygen
bond-stretching vibrations.

Such an enhancement of softening, related to the delocalized components of the electronic state at the Fermi energy, also has been obtained in our calculations of the generic
anomalous OBSM in the cuprate HTSC's. In this case we have shown that with increased hole doping the inclusion of the delocalized Cu4s state into the orbital basis is an important factor for the strength of the
anomalies in the optimally and overdoped state \cite{Ref02,Ref06,Ref20}. Thus, doping (similar like pressure) causes a progressive modification of the orbital character of the electronic state
in the cuprate-HTSC's. Quite generally, our investigations in the cuprates point to an interrelation between electronic structure changes at $\varepsilon_{F}$, selective phonon
softening via strong nonlocal EPI of CF' type and, finally, to the appearance of phonon-mediated superconductivity in an ''ionic'' metal. A corresponding connection seems also to hold in doped Ba-Bi-O.

A change of the electronic state in Ba-Bi-O, i.\,e. a partial shift of the spectral weight from Bi6s to Bi6p, as suggested from our calculations of the OBSM, obviously
occurs parallel with K-doping because the OBSM anomalies are not present in the weakly doped material \cite{Ref08}.
It has been argued in Ref. \onlinecite{Ref25} that K-doping creates localized holes on the O2p orbitals, rather than on Bi atoms. One might speculate, that an
increased population of the Bi6p states as the most delocalized ones lowers the energy by an enhanced hybridization with the O2p hole states making the latter itinerant at a
certain doping level in the metallic state.

An increasing Bi6p character of the electronic state at $\varepsilon_{F}$ also should lead from a more rounded Fermi surface to a more distorted anisotropic shape. Such a change of the geometry of the Fermi surface
would of course be mapped to the "phase space factor" (nesting function) and also to the matrix elements of the electron-phonon spectral function $\alpha^{2}F(\omega)$
and, finally, to the superconducting $T_{c}$ in a phonon-mediated mechanism.

Altogether from our results a multi-orbital model involving CF's, DF's and the corresponding interactions $\widetilde{V}$ in the charge response, Eqs. \eqref{Eq15}, \eqref{Eq16}, between
Bi6s-, Bi6p- and O2p states is needed to obtain a reliable description of the phonon dispersion in an "ionic" metal like Ba-Bi-O, as shown in Fig. \ref{fig04}.

\begin{figure*}
\includegraphics[]{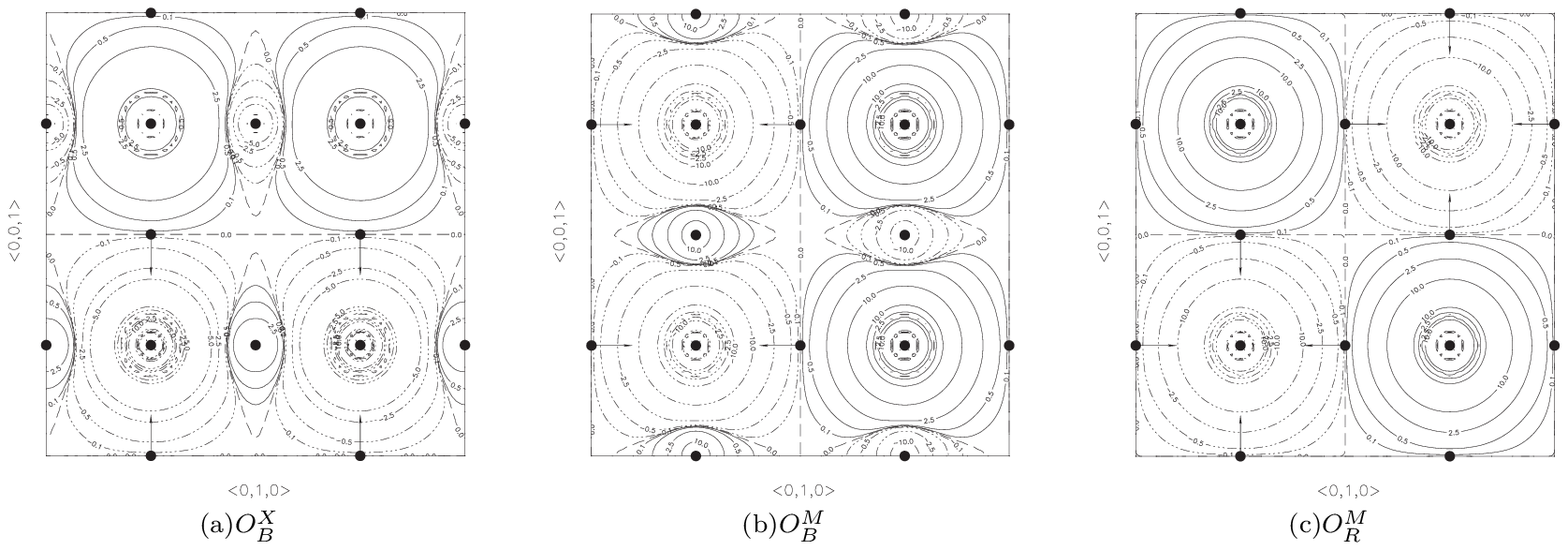}
\caption{Contour plot of the nonlocal part of the phonon-induced charge density redistribution from Eq. \eqref{Eq27} for (a) the $O^{X}_{B}$ mode, (b) the $O^{M}_{B}$ mode and (c) the $O^{R}_{B}$ mode in model 3. The units are $10^{-4}/a_{B}^{3}$. Full lines
($\text{---}$) describe regions of space where electrons are attracted and the lines ($\text{---} \cdots$) represent regions where electrons are pushed away.}
\label{fig07}
\end{figure*}

\begin{figure}
\includegraphics[]{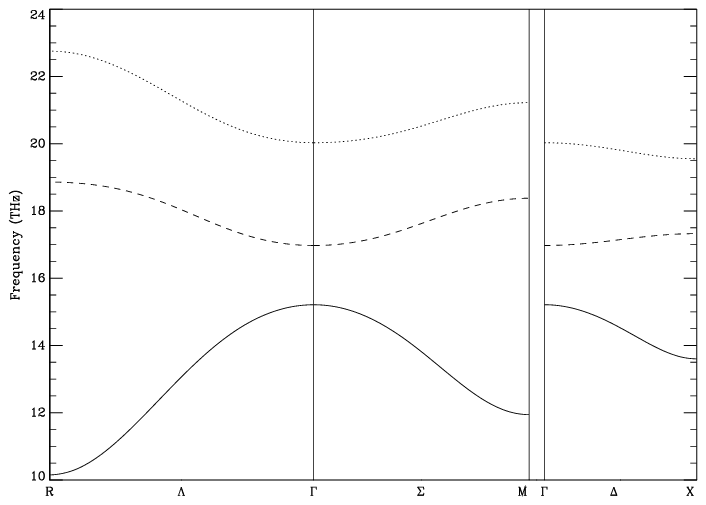}
\caption{Calculated anomalous branches with the oxygen-breathing modes  $O^{X}_{B}$, $O^{M}_{B}$ and $O^{R}_{B}$ as end points of the
$\Delta \sim (0,\,0,\,1)$, $\Sigma \sim (1,\,1,\,0)$ and $\Lambda \sim (1,\,1,\,1)$ direction, respectively. The linetype ($\cdots$) denotes the result for the
RIM (model 1), ($---$) additionally includes DF's (model 2) and the full lines represent the final calculation including DF's and CF's (model 3).}
\label{fig08}
\end{figure}

Comparing the results of the dispersion in Fig. \ref{fig03} and \ref{fig04} we
recognize that the LO-TO splittings at $\Gamma$ are closed by the metallic screening due to the CF's. Such a screening of the longwavelength polar modes does not imply charge
localization on a microscopic scale and can also be achieved in a homogeneous electron gas model. However, localization of charge on a microscopic scale is necessary for the screening of the zone-
boundary modes like $O^{X}_{B}$, $O^{M}_{B}$ and $O^{R}_{B}$. In this case nonlocally induced ionic Bi6s- and Bi6p CF's lead to the strong softening. Compare with Fig. \ref{fig07}, where the
nonlocal part of the phonon-induced charge density redistribution is shown for $O^{X}_{B}$, $O_{B}^{M}$ and $O^{R}_{B}$. The related strong softening of the modes by the nonlocally excited nonrigid
charge response via CF's and DF's as compared to the RIM can be extracted from Fig. \ref{fig08}.

In context with the vanishing of the LO-TO splitting characteristic phonon branches along $\Delta$, $\Sigma$ and $\Lambda$ with the (stable) ferroelectric TO mode at
$\Gamma$ (6.744 THz) appear in the phonon dispersion which are well described in our model. Even details, like the steep slopes near $\Gamma$, resulting from the depression
of the LO frequency by metallic screening, and the flattening out of the branch in $\Lambda$ direction is obtained in our calculation.

\begin{figure*}
\includegraphics[]{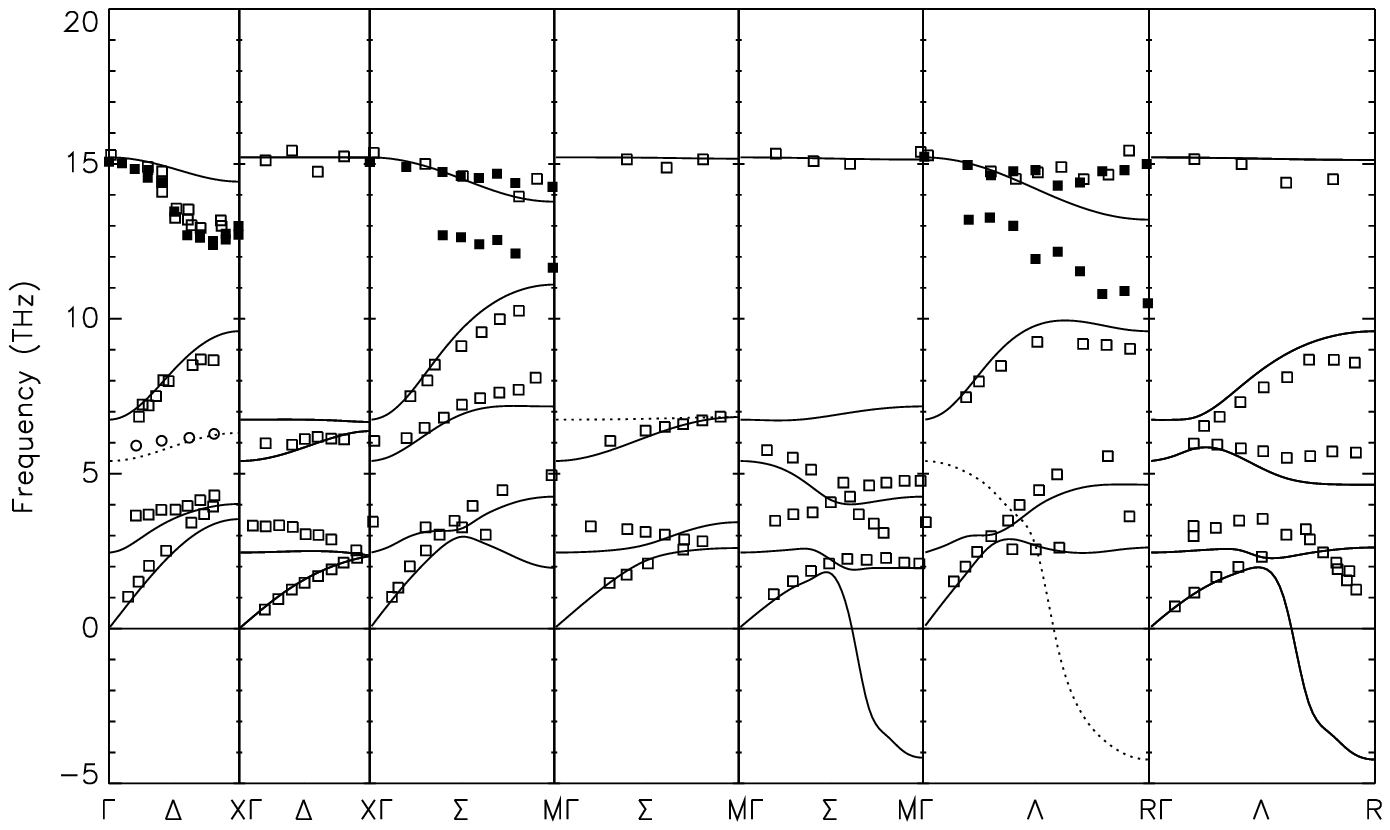}
\caption{Same as in Fig. \ref{fig01} with the calculated results
from the model described in the text where additional to the RIM
dipole fluctuations and charge fluctuations are included. The
latter are based on a model where an electronic density of states
at $\varepsilon_{F}$ typical for LDA calculations of Ba-Bi-O
\cite{Ref24} is used (model 4).} \label{fig09}
\end{figure*}

In order to investigate the question how the phonon dispersion looks like, if we use an electronic density of states at $\varepsilon_{F}$ for Ba-Bi-O which is typical for a LDA-like
calculation \cite{Ref24} in our model for the polarizability, we have performed the calculations shown in Fig. \ref{fig09}. The total density of states from Ref. \onlinecite{Ref24} is 0.8 eV$^{-1}$. Taking for
$\Pi(\text{O2p})$ the same value as in model 3 and using the same ratio for $\Pi(\text{Bi6s})$ and $\Pi(\text{Bi6p})$ we obtain: $\Pi(\text{Bi6s}) = 0.4\,\text{eV}^{-1}$,
$\Pi(\text{Bi6p}) = 0.2\,\text{eV}^{-1}$, $\Pi(\text{O2p}) = 0.2\,\text{eV}^{-1}$ (model 4). Comparing the results for model 4 in Fig. \ref{fig09} with those in Fig. \ref{fig04} we find that the softening
of the oxygen-breathing mode anomalies is only poorly developed, consistent with the ab initio calculations based on DFT-LDA in Ref. \onlinecite{Ref11}. This indicates that the ab initio LDA calculations
underestimate the EPI for the OBSM.

\begin{table}
\begin{tabular}{|c||ccc|}
\hline & $\delta\zeta_{\text{Bi6s}}$ & $\delta\zeta_{\text{Bi6p}}$ & $\delta\zeta_{\text{O2p}}$\\ \hline\hline
$O_{B}^{X}$  & 39.369 & 27.534 & -4.169\\
                         & 32.333 & 18.285 & -1.764\\ \hline
$O_{B}^{M}$ & 61.329 & 42.740 & -5.791\\
                         & 46.309 & 26.227 & -2.092\\ \hline
$O_{B}^{R}$  & 83.657 & 57.859 & / \\
                         & 56.807 & 32.220 & / \\ \hline
\end{tabular}
\caption{Calculated phonon-induced charge fluctuation $\delta\zeta_{\kappa}(\vc{q}\,\sigma)$ for model 3 (first row) and model 4 (second row) for
$O_{B}^{X}$, $O_{B}^{M}$ and $O_{B}^{R}$ according to Eq. \eqref{Eq18}. Units are in $10^{-3}$ electrons. $\delta\zeta_{\kappa}(\vc{q}\,\sigma) > 0$ means that
electrons are pushed away. $/$ means, that no CF's are excited.}\label{tab01}
\end{table}

The underestimation of the EPI in the breathing modes can also be
deduced from Table \ref{tab01} in terms of the CF's. Here the CF's
according to Eq. \eqref{Eq18} of the Bi6s-, Bi6p- and O2p orbitals
generated by $O^{X}_{B}$, $O_{B}^{M}$ and $O^{R}_{B}$ are listed.
We recognize that in our modeling of the electronic state with
enhanced PDOS for Bi6s and Bi6p at $\varepsilon_{F}$ as compared
to the LDA-based result the CF's are significantly larger leading
to the correct description of the anomalies in the phonon
dispersion.

The strong nonlocal EPI of ionic CF type in the oxygen-breathing
modes is also illustrated in the contour plots of the nonlocal
part of the phonon-induced charge density redistribution
$\delta\rho(\vc{r},\,\vc{q}\,\sigma)$ for the OBSM in Fig.
\ref{fig07}, i.\,e.

\begin{equation}\label{Eq27}
\delta\rho(\vc{r},\vc{q}\sigma) = \sum_{\vc{a},\kappa}
\delta\zeta^\vc{a}_\kappa(\vc{q}\sigma)\rho_\kappa(\vc{r}-\vc{R}^\vc{a}_\kappa),
\end{equation}

with the CF's $\delta\zeta^{\vc{a}}_{\kappa}$ according to Eq.
\eqref{Eq18} and the form factors $\rho_{\kappa}$ for the CF's as
in Eq. \eqref{Eq1}. The total phonon-induced charge redistribution
can be obtained by adding the rigid displacement of the
unperturbed densities, i.\,e.,

\begin{equation}\label{Eq28}
\delta\rho_\text{r}(\vc{r},\vc{q}\sigma) = \sum_{\vc{a},\alpha}
\{\rho^0_\alpha(\vc{r}-[\vc{R}^\vc{a}_\alpha+\vc{u}^\vc{a}_\alpha(\vc{q}\sigma)])
-\rho^0_\alpha(\vc{r}-\vc{R}^\vc{a}_\alpha)\}.
\end{equation}

$\rho_{\alpha}^{0}$ is the density of the unperturbed ion from Eq.
\eqref{Eq1} and $\vc{u}^{\vc{a}}_{\alpha}$ the displacement of an
ion in Eq. \eqref{Eq19}.

The results shown in Fig. \ref{fig07} demonstrate that the
nonlocal EPI leads to attractive (-----) and repulsive (-----
$\cdot\cdot$) regions for electrons mainly at the Bi ions. For
symmetry reasons some charge is also transferred to the silent
oxygens in the $O^{X}_{B}$ and $O^{M}_{B}$ mode, see Fig.
\ref{fig07}a, \ref{fig07}b and Table \ref{tab01}. This, however,
is not possible in $O^{R}_{B}$ because all the oxygen ions are
vibrating symmetrically towards or away from Bi. Electrons, i.\,e.
negative charge, is flowing to the expanded octahedra from the
compressed ones. Thus, negative charge is always accumulated in
the larger octahedra and so a (dynamical) charge
disproportionation of the Bi ions is generated.

As mentioned in Ref. \onlinecite{Ref20}, if the intersite charge
transfer in terms of CF's induced nonlocally by the OBSM would
additionally be spin polarized, as in the undoped
antiferromagnetic cuprates or the weakly doped cuprates with
antiferromagnetic correlations of sufficient correlation length,
corresponding spin fluctuations are generated (spin-phonon
interaction via charge transfer). In this way the phonon dynamics
and the CF's change the spin dynamics and vice versa. Such a
spin-phonon coupling may be quite strong and could lead to a
synergetic pairing mechanism in which both phonons and spin
excitations participate, see e.g. Ref. \onlinecite{Ref26} and
references therein. For a qualitative discussion of a possible
interplay between lattice-, charge- and spin degrees of freedom in
the cuprates with the help of the OBSM see also Ref.
\onlinecite{Ref20}.

The CF's of the Bi6s-, Bi6p- and O2p orbitals
$\delta\zeta_{\kappa}(\vc{q}\,\sigma)$ for $O^{X}_{B}$,
$O^{M}_{B}$ and $O^{R}_{B}$ are collected in Table \ref{tab01}.
Significantly larger values are found for model 3 compared with
the LDA-based model 4. For example in the $O^{R}_{B}$ mode 0.142
electrons are exchanged between the vibrating octahedra in model 3
and only 0.089 electrons in model 4.

For lightly K-doped insulating Ba-Bi-O besides the tilt distortion
related to the rotational modes there is also a breathing
distortion (frozen-in breathing mode) of the octahedra. On the
other hand, our calculations in model 3 for the metallic phase
show that strong nonlocal EPI of CF-type lead to a large softening
but a still stable $O_{B}^{R}$ breathing mode in very good
agreement with experiment. However, the result indicates that this
type of nonlocal EPI also should play a role for the actual
breathing distortion in the insulating part of the phase diagram.
One can interpret the $\delta\rho$-patterns in Fig. \ref{fig07}
and the CF's in Table \ref{tab01} as an expression  of a precursor
effect for polaron formation. Actually, polarons are not expected
to form at a high enough doping level, because, as mentioned
before, the system can gain sufficient energy by delocalization
via the hybridization of the Bi6s and in particular the Bi6p
states with the O2p hole states created by doping. However, such a
channel for a lowering of the energy by delocalization might not
be favourable in the insulating, lightly doped material due to the
small number of holes and the tilted structure where additionally
the O2p to Bi6p hopping is decreased due to misalignment of the
orbitals and hence the energy gain by delocalization is reduced.
In this case (polaronic) charge localization by the strong
nonlocal EPI of CF type might be energetically more favourable and
responsible for the metal-insulator transition.

A tendency to polaron formation is also very likely in the undoped
and weakly  doped cuprate-HTSC's, see e.g. Refs.
\onlinecite{Ref27,Ref28}. In this context, our modeling of the
charge response has shown that the delocalized component of the
electronic state, i.\,e., in this case Cu4s, is not required to
describe the phonon anomalies in these "phases" of the cuprates
\cite{Ref02,Ref03,Ref06,Ref20}. Polaronic effects also occur in
the small non-adiabatic region around the $c$-axis of the
cuprat-HTSC's where an insulator-like charge response and
phonon-plasmon mixing is found \cite{Ref20,Ref29}.

Our calculations for La-Cu-O of the orbital resolved phonon induced
change of the self consistent potential $\delta V_\kappa$ felt by an
electron, according to Eq. \eqref{Eq21}, point to a hierarchy of energy
scales for coupling of the electrons with the
phonons\cite{Ref04,Ref20}. There is one energy scale related to the
OBSM with $\delta V_\text{Cu3d}$ up to about 100meV for the
O$_\text{B}^X$ mode (with a frequency of 79meV) and up to about 75meV
for the half breathing mode (with frequency of 69meV). Both modes have
been calculated in adiabatic approximation which seems sufficient for
the latter modes\cite{Ref20}. Moreover, we find in La-Cu-O due to weak
screening perpendicular to the Cu-O plane in a small $\vc{q}$ space
region around the $(0, 0, 1)$ direction in particular in the lightly
doped materials a strong nonlocal, non-adiabatic electron-phonon
coupling of CF type to the charge carriers in the Cu-O plane, which are
responsible for superconductivity in the cuprates, via axial breathing
modes of the apex oxygen and the La ion. The corresponding values of
$\delta V_\kappa$ for these modes increase up to about 800meV near the
phonon-plasmon resonance at low frequencies (phonon-like mode, possibly
overdamped)\cite{Ref20}. Moreover, there is strong coupling at high
frequencies beyond the phonon spectrum for the plasmon-like mode. Thus,
a hierarchy of interaction scales at low energy as well as high energy
is provided by these nonlocal electron phonon coupling effects caused
by weak screening. Quite recently besides the well known low energy
''kink'' scale (40-80meV), see e.g. Refs. \onlinecite{Ref30,Ref31},
also a high energy anomaly in the electron spectrum of HTSC's
(300-800meV) has been observed by angle-resolved photoemission
spectroscopy\cite{Ref32}. See also Refs. \onlinecite{Ref33,Ref34} for
the observation of the high energy anomalies and an anomalous
enhancement of the width of the LDA-based CuO$_2$-band extending to
energies of about 2eV.\cite{Ref34}

As a matter of fact, the electronic structure of the lightly K-doped
insulating phase of Ba-Bi-O is not really understood. However, as
discussed e.g. in Refs. \onlinecite{Ref35,Ref36} and from our
discussion formation of small polaron or bipolarons due to strong
nonlocal EPI may be an important mechanism for insulating behaviour
persisting over a wide doping range.

The mode splitting experimentally found in recent experiments for the
anomalous branches along $\Sigma$ and $\Lambda$ for metallic
Ba-Bi-O\cite{Ref09} (see the full symbols in the figures of the phonon
dispersion) cannot be understood in calculations assuming an ideal
cubic perovskite symmetry. As discussed in Ref. \onlinecite{Ref09} the
split dispersion suggests some charge inhomogeneity due to distortions
from the cubic perovskite structure on a time scale larger than the
time scale of the phonons. For example a structural tilt distortion
around the cubic (0, 0, 1) direction recently reported for
superconducting Ba-Bi-O \cite{Ref37} should lead to an increase of the
OBSM at the $M$- and $R$-point but not at the $X$-point, as seen in the
experiments, because of the decreased O2p to Bi6p hopping due to
misalignement of the orbitals and a corresponding reduction of the
softening effect via Bi6p.

Anomalous phonon softening and mode splitting of the oxygen
bond-stretching phonon along (1, 0, 0), most likely due to a (dynamic)
charge inhomogeneity, very recently also has been observed in
La-Ba-Cu-O \cite{Ref38}. As discussed in Ref. \onlinecite{Ref06} a
charge inhomogeneity with lower symmetry as the ''average''
translational invariant charge inhomogeneity considered in our modeling
of the electronic state of the underdoped
cuprate-HTSC's\cite{Ref06,Ref20} should lead locally to a variation of
the strength of the nonlocal EPI of CF type and a related variation of
the phonon frequency of the OBSM. The nonlocal EPI may become locally
very strong in case of a metallic, compressible charge inhomogeneity.

In our modelling of the electronic state of the underdoped
cuprates we have incompressible, insulator-like regions related to
the localized Cu3d states interpenetrating a metallic,
compressible structure provided by the more delocalized O2p states
at the O$_{xy}$ sublattices, where the holes accumulate. Such a
correlated percolated metallic structure can be thought
essentially to be due to the interplay of the short range
repulsion ($U_d$) of the Cu3d states and the long range Coulomb
interaction together with a gain in kinetic energy of the holes by
delocalization via the O2p states. For higher doping in the
optimal and overdoped state the delocalized Cu4s component of the
electronic state becomes important in our modelling leading to the
experimentally observed strong increase of phonon softening of the
OBSM in the cuprates, see e.g. Ref.\onlinecite{Ref06}. Physically
this means that the percolated structure melts with increased
doping because of the enhanced quantum mechanic tunneling via the
delocalized Cu4s and O2p component in the electronic state at the
the Fermi energy. Ultimately, in the overdoped ''phase'' where in
our modelling the Cu4s component is further enhanced the system
should converge more and more towards a Fermi liquid.

The OBSM are a sensible probe for charge inhomogeneities in form
of regions with a different doping level\cite{Ref06}. How the
frequency would change depends in particular on the electronic
state of the inhomogeneity which could be a mixture of
compressible, metallic (higher doping), incompressible, insulating
(no doping) or partially incompressible (lower doping) regions in
the crystal. The incompressible and partially incompressible
regions (with OBSMs of higher frequencies) most likely are
stabilzed by an interplay of strong electronic correlation and
strong nonlocal EPI of CF type triggering polaronic behaviour,
while in the compressible regions (with lower frequencies of the
OBSMs) delocalization of the electronic structure changes the
nature of the electronic state as discussed in Refs.
\onlinecite{Ref06,Ref20}.

In a phonon-mediated pairing mechanism such a local variation of the
nonlocal EPI by inhomogeneity is expected to induce simultaneously a
corresponding change of the pairing strength (gap value) and a
variation of the frequency of the pairing related modes. Usually,
stronger renormalized modes (lower frequencies), i.e., a stronger
nonlocal EPI, correlates to a larger pairing strength (higher gap
value), and vice versa. Quite recently such a correlation has been
observed locally in scanning tunneling experiments in
Bi$_2$Sr$_2$CaCu$_2$O$_8$ \cite{Ref39} and as one possibility
attributed to electronic-lattice interactions releated to a competing
electronic ordered state (see, e.g. Ref. \onlinecite{Ref38}).

\begin{figure*}
\includegraphics[]{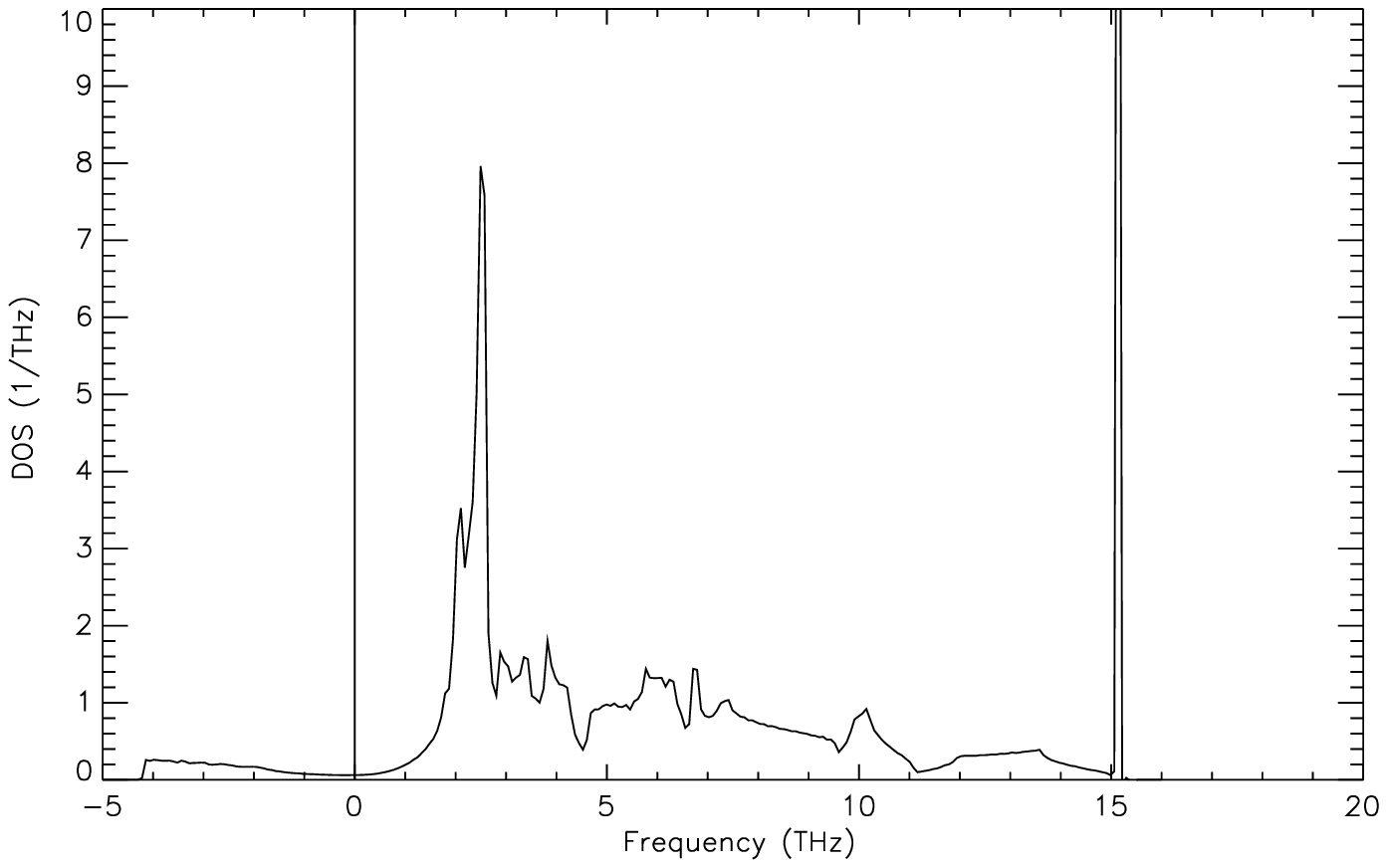}
\caption{Phonon density of states (DOS) for cubic Ba-Bi-O as
calculated with model 3, including CF's and DF's.} \label{fig10}
\end{figure*}

\begin{figure*}
\includegraphics[]{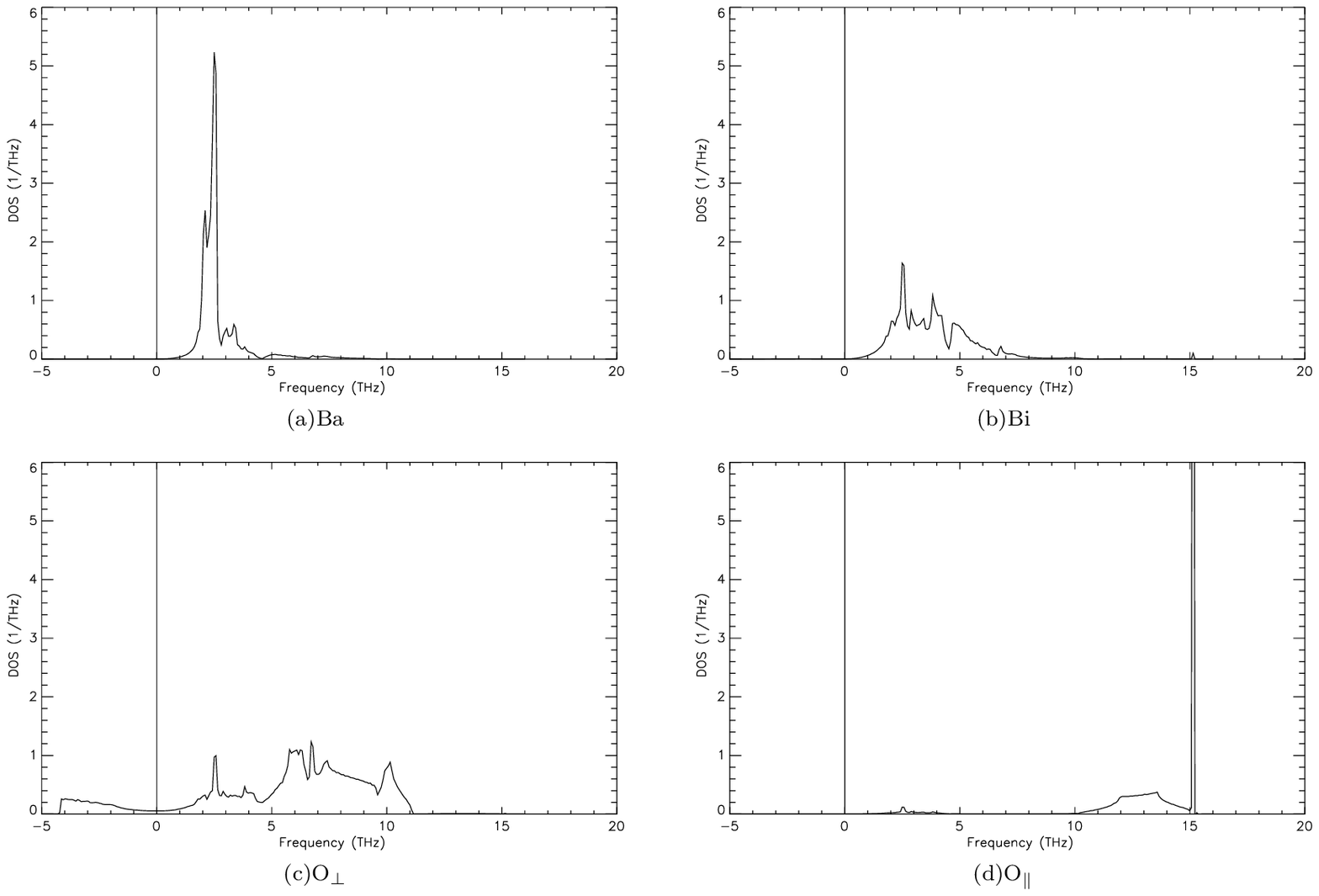}
\caption{Partial phonon density of states calulated with model 3.
Ba-DOS (a); Bi-DOS (b); O$_{\perp}$-DOS (c); O$_{\|}$-DOS (d). The
symbol $\|$ or $\perp$, respectively, means parallel or
perpendicular to the Bi-O bond.} \label{fig11}
\end{figure*}

In the last topic of this paper we present calculations for the
phonon density of states within model 3 including CF's and DF's,
see Fig. \ref{fig10}. Moreover, in Fig. \ref{fig11} the calculated
atom-resolved phonon density of states is shown. Concerning the
oxygen ion we have further discriminated between displacements
parallel (O$_{\|}$) and perpendicular (O$_{\perp}$) with respect
to the Bi-O bond in order to demonstrate the strongly anisotropic
behaviour of the oxygen vibrations in Ba-Bi-O.

From the figures we see that the frequency range up to about 5 THz
can be attributed to Ba and Bi vibrations with some contribution
from O$_{\perp}$ including the unstable rotational modes. The
O$_{\perp}$ vibrations dominate the spectral range from around 5
THz to about 11 THz. Beyond this frequency range the
high-frequency part from 10-15 THz is exclusively governed by the
O$_{\|}$ vibrations including the von Hove points of the OBSM:
$O^{X}_{B} \approx 13.6\,\text{THz}$, $O^{M}_{B} \approx
11.9\,\text{THz}$ and $O^{R}_{B} \approx 10.2\,\text{THz}$. The
anisotropy between the O$_{\|}$ and O$_{\perp}$ vibrations will
for example be reflected in the Debye-Waller factors. The latter
are expected to be considerably larger perpendicular to the Bi-O
bond in agreement with experiments \cite{Ref10} and consistent
with a tendency towards a structural phase transition via
rotational modes. Finally, the strong almost singular peak at
about 15 THz results from the characteristic nearly dispersionless
branches in Fig. \ref{fig04} terminating at the $\Gamma$ point in
the TO mode with the highest frequency.

\section{Summary and discussion}
Within our microscopic modeling of the electronic charge response
in linear response theory we have studied complete phonon
dispersion curves, phonon-induced charge density redistributions
and charge fluctuations, generated by nonlocal EPI, and the total
and atom-resolved phonon density of states in cubic metallic
Ba-Bi-O. Our calculated results of the phonon dispersion curves
are in a good overall agreement with the experiments, in
particular the strong phonon softening of the anomalous
oxygen-breathing modes is well described. This was not possible so
far neither with empirical nor with ab initio calculations based
on the LDA using a virtual crystal approximation. The latter seem
to underestimate the kinetic single particle part of the density
response. Increasing the electronic PDOS for the Bi6s and Bi6p
orbitals at the Fermi level as compared with typical values
obtained in LDA calculations has been shown to provide  the strong
softening of the OBSM. This is achieved by a characteristic type
of local screening on microscopic scale in terms of ionic CF's
being quite generally at work in ''ionic'' metals with a
significant component of ionic binding like Ba-Bi-O or the the
cuprate based HTSC's. More explicitly, nonlocal coupling of the
displaced oxygen ions in the OBSM essentially to Bi6s- and Bi6p
CF's, which are favoured by the corresponding large PDOS of these
states at $\varepsilon_{F}$, are found to be responsible for the
softening.

The effect of the strong nonlocal EPI is also visible from the
calculation of the nonlocal part of the charge density
redistribution and the CF's themselves during the oxygen
vibrations in the octahedra. Accordingly, a considerable
(dynamical) charge disproportionation has been calculated. The
patterns of charge rearrangement can be looked upon as a
visualization of a precursor effect for polaron formation. Small
polaron and/or bipolaron formation is very likely an important
mechanism for the insulating behaviour in lightly K-doped Ba-Bi-O.

The strong effect for softening by including the Bi6p orbitals
into the electronic state at the Fermi energy points towards a
change of this state with K-doping. This change is accompanied by
a corresponding variation of the topology of the Fermi surface and
possible changes of the electron-phonon matrix elements. Both
effects are important for the electron-phonon spectral function
$\alpha^{2}F(\omega)$ and the $T_{c}$ value in a phonon-mediated
mechanism for superconductivity. Such a mechanism is supported by
the strong nonlocal EPI effects found in our calculations.
Compared with the cuprate based HTSC spin degrees of freedom seem
to play no role for superconductivity in Ba-Bi-O but the strong
nonlocal mode-selective coupling effects of lattice and charge
degrees of freedom in terms of ionic CF's are present in both,
Ba-Bi-O and the cuprate-HTSC's.

From our calculations of the PDOS of the phonons we extract that
the oxygen vibrations and consequently also the mean square
displacements are very anisotropic. There are low-frequency
O$_{\perp}$ vibrations perpendicular to the Bi-O bond which are
well separated from the high-frequency O$_{\|}$ vibrations
parallel to the bond.

Quite generally, our investigations for metallic Ba-Bi-O and the
cuprate-HTSC's point to an interrelation between doping induced
electronic structure changes at the Fermi energy, selective phonon
softening via strong nonlocal EPI of CF' type and high-temperature
superconductivity in "ionic" metals.

\end{document}